\documentclass[twoside]{article}
\usepackage{fleqn,espcrc2}

\usepackage{graphicx}
\usepackage[figuresright]{rotating}

\newcommand{\AmS}{{\protect\the\textfont2
  A\kern-.1667em\lower.5ex\hbox{M}\kern-.125emS}}

\title{Simulations of Protein Folding}

\author{Michael Cahill,\address{Department of Chemistry, 
        United States Military Academy,
        West Point, NY 10997, USA}
        Mark Fleharty,\address{Department of Computer Science,
	University of New Mexico,
	Albuquerque, NM 87131, USA}
        and 
        Kevin Cahill\address{Department of Physics and Astronomy,
        University of New Mexico,
        Albuquerque, NM 87131, USA}}
       
\begin{document}

\begin{abstract}
We have developed a simple, phenomenological, 
Monte-Carlo code that predicts the three-dimensional structure
of globular proteins from the DNA sequences that define them.
We have applied this code to two small proteins,
the villin headpiece (1VII) and cole1 rop (1ROP)\@.
Our code folds both proteins to within 5 \AA\ rms of 
their native structures.
\vspace{1pc}
\end{abstract}

\maketitle

\section{PROTEINS}
A protein is a linear chain of amino acids.
The proteins of natural living organisms
are composed of 20 different types of amino acids.
A typical protein is a polymer of 300
amino acids, of which there are
\(
20^{300} = 2 \times 10^{390}
\)
different possibilities.
The human body uses about 80,000 different proteins
for most of its functionality,
including structure, communication, transport,
and catalysis.  

The order of the amino acids in the proteins
of an organism
is specified by the order of the base pairs
in the deoxyribonucleic acid, DNA, of its genome.
Human DNA consists of $10^9$ base pairs
with a total length of 3m.
Since three base pairs specify an amino acid,
the code for the 80,000 human proteins requires
only
\(
3 \times 300 \times 80,000 = 7 \times 10^7
\)
base pairs or 7\% of the genome.
\subsection{Amino Acids}
The twenty amino acids differ only in their side chains.
The key atom in an amino acid
is a carbon atom called the $\alpha$-carbon, C$_\alpha$.
Four atoms are attached to the C$_\alpha$
by single covalent bonds:
a hydrogen atom H,
a carbonyl-carbon atom C$'$,
a nitrogen atom N,
and the first atom of the side chain R
of the amino acid.
The carbonyl carbon C$'$ is connected
to an oxygen atom by two covalent bonds
and to a hydroxyl group OH by another covalent bond;
the nitrogen atom N is attached to
two hydrogen atoms, forming an amine group NH$_2$.
The backbone of an amino acid is the triplet
N, C$_\alpha$, C$'$.

Of the 20 amino acids
found in biological systems, 19 
are left handed.
If one looks at the C$_\alpha$ from the
H, then the order of the structures C$'$O,
R, and N is clockwise (CORN).
The one exception is glycine in which
the entire side chain is a single hydrogen atom;
glycine is not chiral.

\subsection{Globular Proteins}
There are three classes of proteins:
fibrous, membrane, and globular.
Fibrous proteins are the building materials of bodies;
collagen is used in tendon and bone,
$\alpha$-keratin in hair and skin.
Membrane proteins sit in the membranes of cells
through which they pass molecules and messages.
Globular proteins catalyze chemical reactions;
enzymes are globular proteins.
                                                       
Under normal physiological conditions,
saline water near pH=7 at 20--40 \(^{\rm o}\)C,
proteins assume their native forms.
Globular proteins fold into compact structures.
The biological activity of a globular protein
is largely determined by its unique shape,
which in turn is determined by its primary structure,
that is, by its sequence of amino acids.
\subsection{Kinds of Amino Acids}
The amino acids that occur in natural living
organisms are of four kinds.
Seven are nonpolar: alanine (ala), valine (val), phenylalanine (phe),
proline (pro), methionine (met),
isoleucine (ile), and leucine (leu).
They avoid water and are said to be \emph{hydrophobic}.
Four are charged: aspartic acid (asp) and glutamic acid (glu)
are negative,
lysine (lys) and arginine (arg) are positive.
Eight are polar:  serine (ser), threonine (thr),
tyrosine (tyr), histidine (his),
cysteine (cys), asparagine (asn), glutamine (gln),
and tryptophan (trp).
The four charged amino acids 
and the eight polar amino acids seek water
and are said to be \emph{hydrophilic}.
Glycine falls into a class of its own.          
\subsection{Protein Geometry}
When two amino acids are joined to make
a dipeptide, first the hydroxyl group OH attached
to the carbonyl carbon C$'$ of the first amino acid
combines with one of the two hydrogen atoms
attached to the nitrogen N of the second amino acid
to form a molecule of water H$_2$O, and then a peptide bond
forms between the carbonyl carbon C$'$ of the first amino acid
and the nitrogen N of the second amino acid.
A peptide bond is short, 1.33 \AA, and resists rotations
because it is partly a double bond.

To a good approximation, the six atoms C$_{\alpha 1}$,
C$'_1$, O$_1$, N$_1$, H$_1$, and C$_{\alpha 2}$ lie in a
plane, called the peptide plane.  If a third amino acid
is added to the carbonyl carbon C$'_2$ of the second amino acid,
then the six atoms C$_{\alpha 2} \dots$C$_{\alpha 3}$
also will lie in a (typically different) plane.
Exceptionally, the peptide plane of proline
is not quite flat because
the side chain loops around, and
its third carbon atom
forms a bond with the nitrogen atom
of the proline backbone.   
\subsection{The Protein Backbone}
The protein backbone consists of the chain of triplets
(N C$_\alpha$ C$'$)$_1$,
(N C$_\alpha$ C$'$)$_2$, (N C$_\alpha$ C$'$)$_3$, $\dots$,
(N C$_\alpha$ C$'$)$_N$.
Apart from the first nitrogen N$_1$
and the last carbonyl carbon C$'_n$,
this backbone (and its oxygen
and amide hydrogen atoms)
consists of a chain of peptide planes,
C$_{\alpha 1} \dots$C$_{\alpha 2} \dots$
C$_{\alpha n-1} \dots$C$_{\alpha n}$.
Since the angles among the four bonds of the
C$_\alpha$'s are fixed, the shape of the backbone
of peptide planes is determined by the angles of rotation
about the single bonds that link each C$_{\alpha}$
to the N that precedes it and the C$'$ that follows it.
The angle about the N$_i$-C$_{\alpha i}$ bond is
called $\phi_i$, that about the C$_{\alpha i}$-C$'_i$ bond
is $\psi_i$.  The 2N angles $(\phi_1,\psi_1) \dots
(\phi_N,\psi_N)$ determine the shape of the backbone 
of the protein.
These angles are the main kinematic variables
of a protein.
The principal properties of proteins
are discussed in the classic article
by Jane Richardson~\cite{Jane}.
\section{PROTEIN FOLDING}
The problem of protein folding
is to predict the natural folded shape
of a protein under physiological conditions
from the DNA that defines 
its sequence of amino acids,
which is its primary structure.
This difficult problem has been approached
by several techniques.  Some scientists have
applied all-atom molecular dynamics~\cite{mdVH}.
We have used the Monte Carlo method
in a manner inspired by the work of Ken Dill \emph{et al.}~\cite{dill}. 

Our Monte Carlo simulations 
are guided by a simple potential with three terms.
The first term embodies the Pauli exclusion principle.
Because the outer parts
of atoms are electrons which are fermions,
the Pauli exclusion principle requires that
the side chains of a protein not overlap by more than
a fraction of an angstrom.
In our present simulations, we have represented
each side chain as a sphere centered at the
first carbon atom, the \(C_\beta\), of the side chain
or at the hydrogen atom that is the side chain
in the case of glycine.

The second term represents the mutual attraction of nonpolar
or hydrophobic amino acids.
In effect the water electric dipoles, the free protons,
the free hydroxyl radicals, and the other ions of the cellular fluid
attract the charged and polar amino acids of a protein 
but leave unaffected the nonpolar amino acids.
The resulting net inward force on the
nonpolar amino acids drives them into a core
which can be as densely packed as an ionic crystal.

The third term is a very phenomenological
representation of the effects of steric
repulsion and hydrogen bonding.
For a given amino acid,
this term is more negative when its pair
of angles $\phi_i$ and $\psi_i$ are in a
zone that avoids steric clashes
between the backbone and the side chain
and that encourages the formation of hydrogen bonds
between NH$^{+}$ and O$^-$ groups.
One of these Ramachandran zones
favors the formation of \(\alpha\) helices,
others favor \(\beta\) structures.     

We incorporate these zones in a Metropolis step
with two scales, which we call zoning with memory.  
Each Monte-Carlo trial move
begins with a random number that determines whether
the angles $\phi_i$ and $\psi_i$ of residue \(i\)
will change zone, \emph{e.g.,} from its present zone
to the \(\alpha\) zone, the \(\beta\) zone, or to the
miscellaneous zone.
If the zone is changed, then the angles $\phi_i$ and $\psi_i$
revert to the values they possessed when residue \(i\) 
was last in that zone.
The trial move is then modified slightly and randomly.

\subsection{Rotations}
We have derived a simple
formula for the 3\(\times\)3 real orthogonal
matrix that represents a right-handed rotation 
by \(\theta = |\vec \theta|\) radians about the
axis \(\hat \theta = \vec \theta/ \theta \):
\[
e^{-i \vec \theta \cdot \vec J} = 
\cos \theta \, I -i \hat \theta \cdot \vec J \, \sin \theta
+ ( 1 - \cos \theta ) \, \hat \theta ( \hat \theta )^\mathsf{T}
\]
in which the generators \( (J_k)_{ij} = i \epsilon_{ikj} \) satisfy
\( [ J_i, J_j ] = i \epsilon_{ijk} J_k \)
and \( \mathsf{T} \) means transpose.
In terms of indices, this formula for
\( R( \vec \theta ) = e^{-i \vec \theta \cdot \vec J} \) is
\[
R( \vec \theta )_{ij} =
\delta_{ij} \cos \theta - \sin \theta \, \epsilon_{ijk} \hat \theta_k
+ ( 1 - \cos \theta ) \, \hat \theta_i \hat \theta_j.
\]
In these formulae \( \epsilon_{ijk} \)
is totally antisymmetric with \( \epsilon_{123}=1 \),
and sums over \(k\) from 1 to 3 are understood.

\subsection{Distance}
A conventional measure of the
quality of a theoretical fold
is the mean root-mean-square distance \( d \)
between positions \( \vec r(i) \) 
of the \(\alpha\) carbons
of the folded protein and those \( \vec x(i) \)
of the native structure of the protein,
\[
d = \sqrt{ {1 \over n} \sum_{i=1}^n 
\left( \vec r(i) - \vec x(i) \right)^2 }.
\label{d}
\]
The native states of many proteins are 
available from http://www.rcsb.org/pdb/. 
We have derived a formula for this distance
in terms of the centers of mass 
\( \vec r = (1/n) \sum_{j=1}^n \vec r(j) \)
and \( \vec x = (1/n) \sum_{j=1}^n \vec x(j) \),
the relative coordinates
\( \vec q(i) = \vec r(i) - \vec r \) and
\( \vec y(i) = \vec x(i) - \vec x \),
their inner products \( Q^2 = \sum_{i=1}^n \vec q(i)^2 \)
and \( Y^2 = \sum_{i=1}^n \vec y(i)^2 \),
and the matrix that is the sum of their outer products
\( B = \sum_{i=1}^n \vec q(i) \vec y(i)^\mathsf{T}\). 
If \( (B^\mathsf{T})_{kl} = B_{lk} = \sum_{i=1}^n q(i)_l y(i)_k \) 
denotes the transpose
of this 3\(\times\)3 matrix \(B\)
and tr denotes the trace, then
the rms distance \( d \) is
\[
d = \left\{ {1\over n} \left[ Q^2 + Y^2 
-2 \, \mathrm{tr} \! 
\left(B \, B^\mathsf{T} \right)^{{1\over2}} \right] \right\}^{{1\over2}}.
\]

\subsection{Two Proteins}
We have performed
simulations on a protein fragment of 36 amino acids
called the villin headpiece (1VII)\@.
We begin by rotating the \(2n\) dihedral angles
\(\phi\) and \(\psi\) 
of the protein to \( \pi \),
except for the angle \(\phi\) of proline.
In this denatured starting configuration,
the average rms distance \(d\) is 29 \AA\@.
Our best simulations so far fold the 
villin headpiece 
to a mean rms distance \(d\)
that is slightly less than 5 \AA$\>$
from its native state.
\par
Our second protein is a 56-residue fragment
of the 63-residue protein cole1 rop (1ROP)\@.
From a denatured configuration with \( d = 55\) \AA,
our code folds this protein
to a mean rms distance \(d\) 
of slightly less than 3.2 \AA\ from its native state.

\section*{ACKNOWLEDGEMENTS}
We wish to thank Ken Dill for many
key suggestions;
Charles Beckel, John McIver, Susan Atlas, and Sorin Istrail
for helpful conversations;
Sean Cahill and Gary Herling for critical readings 
of the manuscript;
and Sau Lan Wu and John Ellis
for their hospitality at CERN. 
We have performed our computations on two (dual Pentium II)
personal computers running Linux;
we are grateful to Intel and Red Hat
for reducing the cost of scientific computing.

\end{document}